\documentclass[twocolumn,prb,superscriptaddress]{revtex4}
\usepackage{graphicx}
\usepackage{dcolumn}
\usepackage{bm}
\usepackage{amsmath}
\usepackage{times}
\usepackage{epstopdf}
\usepackage{color}
\usepackage[breaklinks=true,colorlinks,citecolor=blue,linkcolor=blue,urlcolor=blue]{hyperref}
\makeatletter

\newcommand{\Rmnum}[1]{\expandafter\@slowromancap\romannumeral #1@}
\makeatother
\graphicspath{{images/}}

\begin{document}

\author{Mahammad Tahir}
\affiliation{Department of Physics, Indian Institute of Technology Kanpur, Kanpur- 208016, India}
\author{Somya Diwakar}
\affiliation{Department of Physics, Indian Institute of Technology Kanpur, Kanpur- 208016, India}
\author{Subhakanta Das}
\affiliation{Department of Physics, Indian Institute of Technology Kanpur, Kanpur- 208016, India}
\author{Mukul Gupta}
\affiliation{UGC-DAE Consortium for Scientific Research, University Campus, Khandwa Road, Indore- 452001, India}
\author{Rohit Medwal}
\affiliation{Department of Physics, Indian Institute of Technology Kanpur, Kanpur- 208016, India}
\author{Soumik Mukhopadhyay}
\email{soumikm@iitk.ac.in}
\affiliation{Department of Physics, Indian Institute of Technology Kanpur, Kanpur- 208016, India}
\title{Enhanced Spin Pumping and Magnetization dynamics in Ni$_{80}$Fe$_{20}$/MoS$_{2}$ stack via Interface Modification}
\begin{abstract}

Materials with strong spin orbit coupling (SOC) are essential for realizing spin orbit torque (SOT) based magnetic memory devices. Transition metal dichalcogenides (TMDs) are promising candidates for such applications because of their inherently high SOC strength. In this study, we investigate the spin pumping effect at the interface between a monolayer of molybdenum disulfide (ML-$\mathrm{MoS_{2}}$) and $\mathrm{Ni_{80}Fe_{20}}$ (Py) thin films using broadband ferromagnetic resonance (FMR) spectroscopy. FMR measurements reveal a notable enhancement in the effective Gilbert damping factor for the ML-$\mathrm{MoS_{2}/Py}$ (Pt = 0 nm) interface compared to the reference Py thin films, attributed to spin pumping across the ML-$\mathrm{MoS_{2}/Py}$ interface. To further quantify spin pumping efficiency, we introduce a high SOC platinum (Pt) interlayer at the ML-$\mathrm{MoS_{2}/Py}$ interface and systematically vary its thickness. This allows us to evaluate key spin transport parameters, including the enhancement in the effective Gilbert damping parameter, the effective spin mixing conductance that reflects the transfer of spin angular momentum from Py to ML-$\mathrm{MoS_{2}}$ and the effective spin current density.  

\end{abstract}
\maketitle
\section{Introduction}

The key aspects of pure spin current based devices include the generation, transportation, and detection of spin currents, which play a crucial role in spintronic applications. Pure spin currents enable information transfer without charge flow, exert spin transfer torque on magnetic materials, and facilitate spin to charge interconversion~\cite{Maekawa}. Several mechanisms contribute to the generation of pure spin currents, including the spin Hall effect~\cite{Liu,Sinova}, Rashba Edelstein effect~\cite{Wang,Manchon}, spin pumping~\cite{Tser1,Tser2,Tser3}, current injection into a lateral spin valve using a nonlocal geometry~\cite{Fukuma,Demidov}, and spin caloric effects~\cite{Wees}. 

Recently, two dimensional (2D) semiconducting layered transition metal dichalcogenides (TMDs) have garnered significant interest over conventional heavy metals (HMs) due to their unique electronic band structures~\cite{Mak,Zhu}, valley (pseudospin) effects, strong spin orbit coupling (SOC), and broken inversion symmetry in their crystal structures, which enable distinct charge and spin transport phenomena~\cite{Mak1,Cao}. Unlike HMs, where crystallinity control at ultrathin thicknesses is challenging and bulk effects dominate spin orbit torque (SOT), TMDs can generate a pure spin current within a monolayer, apart from bulk contributions. This makes them highly attractive for spintronic applications, especially in energy efficient memory and logic devices. TMDs exhibit 2D electronic states with pronounced spin momentum locking, wherein the charge carriers move in such a way that their momenta are always perpendicular to their spin. This property is crucial for achieving efficient spin to charge interconversion. Structurally, TMDs adopt the general formula $\mathrm{TX_{2}}$, where T represents a transition metal (e.g., Mo, W) and X represents a chalcogen (S, Se, Te)~\cite{Shao,Zadeh}. These materials exhibit a layered hexagonal structure, in which a transition metal atom is sandwiched between two hexagonal planes of chalcogen atoms, coordinated by covalent bonding in a trigonal prismatic configuration~\cite{Mak,Zhu,Shao,Zadeh}. Among various TMDs, monolayer (ML) $\mathrm{MoS_{2}}$ has emerged as a promising material due to its abundance, non toxicity, environmental stability, and highly tunable electronic properties, even at atomic thicknesses~\cite{Mak,Zhu}. Crystallizing in the space group P6m2 (point group D3h), ML-$\mathrm{MoS_{2}}$ exhibits strong SOC induced valence band splitting~\cite{Zhu}. Depending on symmetry considerations, spin splitting can be categorized as Rashba type~\cite{Xiao,Cheng} or Zeeman type~\cite{Yuan}. First principles calculations reveal that the SOC strength in ML-$\mathrm{MoS_{2}}$ reaches approximately 150 meV in the valence band, whereas it is a few tens of meV in the conduction band~\cite{Zhu}. This strong SOC plays a critical role in spin to charge interconversion and interfacial spin transport phenomena.  

The presence of significant SOC in ML-$\mathrm{MoS_{2}}$ facilitates strong interfacial hybridization when coupled to a ferromagnetic (FM) layer~\cite{Shao, Kos, Feng}. Experimental studies have shown that ML-$\mathrm{MoS_{2}}$ acts as an efficient spin sink material when interfaced with various FM layers, such as Co~\cite{Cheng}, $\mathrm{Co_{60}Fe_{20}B_{20}}$~\cite{Gupta}, $\mathrm{Co_2FeAl}$~\cite{Husain}, $\mathrm{Co_2FeSi}$~\cite{Sharma1}, and $\mathrm{Ni_{81}Fe_{19}}$~\cite{Bangar}. The 2D nature of ML-$\mathrm{MoS_{2}}$ suppresses extrinsic contributions to magnetic damping, thereby providing a clean platform to probe intrinsic spin orbit driven phenomena~\cite{Soumyanarayanan, Parkin}.  
Moreover, inversion symmetry breaking in ML-$\mathrm{MoS_{2}}$ significantly enhances SOC effects, leading to strong spin orbit torques (SOTs) driven by mechanisms such as the Rashba Edelstein effect and the spin Hall effect~\cite{Wang,Manchon,Liu,Sinova}. These effects facilitate efficient spin current generation and manipulation, which is advancing low power spintronic devices. ML-$\mathrm{MoS_{2}}$/FM heterostructures show strong potential for next generation spin orbitronic devices, including nonvolatile memory, spin field effect transistors, and logic circuits. Its unique structural, electronic, and spintronic properties make it an ideal platform for SOC driven spin transport and SOT studies. Its strong SOC, spin momentum locking, and FM compatibility make it a promising platform for spintronics and quantum technologies. 

In this study, we systematically investigate the magnetization dynamics and spin pumping efficiency at the $\mathrm{Si/SiO_{2}/ML-MoS_{2}/Ni_{80}Fe_{20}}$ [Py (5 nm)] interface, incorporating a high spin orbit coupling (SOC) material, platinum (Pt), as an interlayer (IL) of varying thickness. Broadband ferromagnetic resonance (FMR) spectroscopy is employed to probe the dynamic magnetic properties and quantify the influence of interfacial engineering on spin transport phenomena. This study highlights the crucial role of ML-$\mathrm{MoS_{2}}$ as an efficient spin sink, leveraging its high crystalline quality and room temperature stability. The integration of ML-$\mathrm{MoS_{2}}$ with Py (5 nm) and a Pt interlayer enhances spin pumping efficiency, making $\mathrm{Si/SiO_{2}/ML-MoS_{2}/Py(5 nm)}$ heterostructures promising candidates for energy efficient spin orbit torque (SOT) based memory and logic applications.

\begin{figure}
\includegraphics[width=0.99\linewidth]{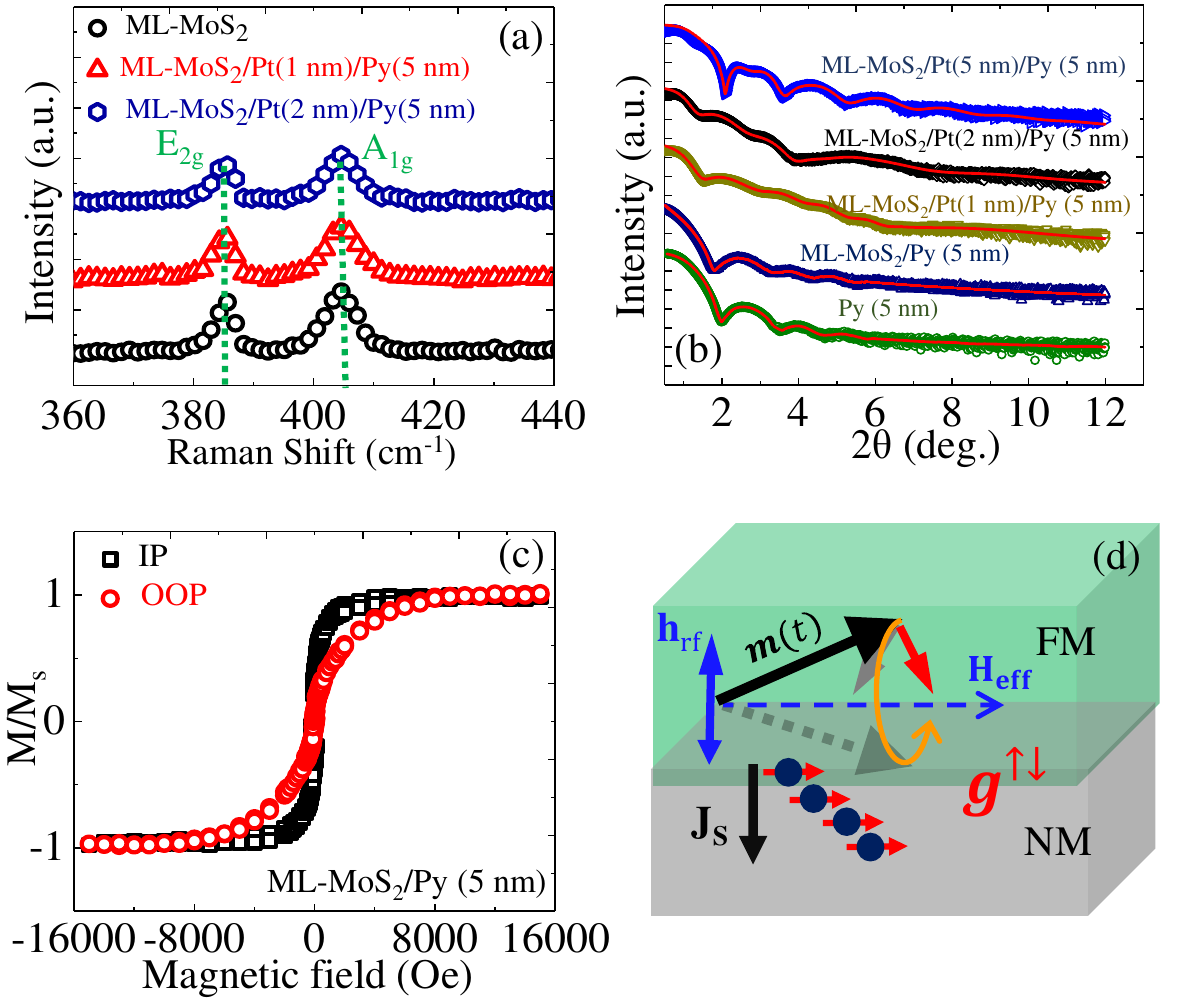}
\caption{(a) Shows the raman spectra of chemical vapor deposition grown ML-$\mathrm{MoS_{2}}$ on a $\mathrm{Si/SiO_{2}}$ substrate and Si/SiO$_{2}$/ML-MoS$_{2}$/Pt(1,2 nm)/Py(5 nm) samples. (b) XRR spectra of Si/SiO$_{2}$/Py(5 nm) and Si/SiO$_{2}$/ML-MoS$_{2}$/Pt(0,1,2,5 nm)/Py(5 nm) interfaces. Open symbols are experimental data points while solid lines represent simulations. (c) In plane (open black squares) and out of plane (open red circles) magnetization hysteresis loops for the Si/SiO$_{2}$/ML-MoS$_{2}$/Py(5 nm) stacks. (d) A schematic of the FM/NM stack is shown to demonstrate how spin pumping facilitates the generation and flow of spin current $\mathrm{J}_{\mathrm{s}}$ across the FM/NM interface.}
\label{fig1}
\end{figure}

\section{Experimental Methods and Characterization}

We prepared the following thin film stacks: Si/SiO$_{2}$/Py(5 nm), Si/SiO$_{2}$/ML-MoS$_{2}$/Py(5 nm) and Si/SiO$_{2}$/ML-MoS$_{2}$/IL/Py(5 nm) with Pt(1,2,5 nm) as an interlayer (IL). We used commercially available chemical vapor deposition (CVD) grown ML (manufactured by OSSILA) on a Si/SiO$_{2}$(300 nm) substrate. We deposited the thin films of IL Pt and a ferromagnetic layer Py by using DC magnetron sputtering on top of ML-MoS$_{2}$ deposited Si/SiO$_{2}$ substrates with a base pressure less than 3 × 10$^{-7}$ Torr. The deposition was carried out at a pressure of 3 × 10$^{-3}$ Torr with an argon flow rate set at 15 standard cubic centimeters per minute (SCCM) to avoid damage to the ML-MoS$_{2}$ layer. The target was pre sputtered for two minutes to avoid contamination during the deposition. The monolayer characteristic of MoS$_{2}$ on the Si/SiO$_{2}$ substrate was characterized by Raman spectroscopy (Model: Princeton Instruments Acton Spectra Pro 2500i) equipped with a 532 nm laser. The Raman spectra are recorded at a laser wavelength of 532 nm with a spot size of 0.5 $\mathrm{\mu}$m. The low laser power of 125 $\mathrm{\mu}$W is used to avoid any heating effect or thermal damage.
\begin{table*}
        \caption{The simulated values of density, thickness and interface roughness of each layer of Si/SiO$_{2}$/ML-MoS$_{2}$/Pt(0,1,2,5 nm)/Ni$_{80}$Fe$_{20}$(5 nm) [Py(5 nm)] stacks are shown }\label{table1}
        \scalebox{0.75}{
    \begin{tabular}{@{\extracolsep{8pt}}lccccccccc@{}} \hline
        \multicolumn{1}{l}{} &
        \multicolumn{3}{c}{Density ( g/cc)} &
        \multicolumn{3}{c}{Thickness (nm)} &
        \multicolumn{3}{c}{Interface Roughnesses (nm)} \\\\
        \cline{2-4}
        \cline{5-7}
        \cline{8-10}

Sample              & ML-MoS$_2$       & Pt   & Ni$_{80}$Fe$_{20}$       & ML-MoS$_2$          & Pt    & Ni$_{80}$Fe$_{20}$        & Substrate/ML-MoS$_2$ & ML-MoS$_2$/Pt  & Ni$_{80}$Fe$_{20}$ \\ 

\hline
            &               &               &                   &           &           &           &       &   &\\

ML-MoS${_2}$/Pt(5 nm)/Ni$_{80}$Fe$_{20}$ (5 nm)   &      5.07$\pm$0.15   &     19.90 $\pm$0.84   &       8.83 $\pm$0.34   &       0.86$\pm$0.04   &      5.03$\pm$0.14  &       5.08$\pm$0.34   &      0.33$\pm$0.02   &        0.34$\pm$0.04   &      0.25$\pm$0.06   \\
   \\        
ML-MoS${_2}$/Pt(2 nm)/Ni$_{80}$Fe$_{20}$ (5 nm)        &       4.98$\pm$0.74  &      20.95$\pm$0.68   &      8.89$\pm$0.68   &       0.89$\pm$0.28&       2.20$\pm$0.02&       5.40$\pm$0.76&      0.28$\pm$0.03 &        0.38$\pm$0.06   &      0.23$\pm$0.07  \\
 \\         
ML-MoS${_2}$/Pt(1 nm)/Ni$_{80}$Fe$_{20}$ (5 nm)        &        4.96 $\pm$0.94   &       21.12 $\pm$0.34   &        8.82$\pm$0.74   &        0.96 $\pm$0.06   &      1.18$\pm$0.26 &       4.88$\pm$0.25   &       0.38 $\pm$0.04  &      0.32$\pm$0.06   &        0.26$\pm$0.04   \\
 \\
ML-MoS${_2}$/Ni$_{80}$Fe$_{20}$ (5 nm)   &        4.98$\pm$0.68     &      0.00$\pm$0.00  &      8.87 $\pm$0.64  &       0.92$\pm$0.03  &      0.00$\pm$0.00   &     5.82$\pm$0.53   &       0.30 $\pm$0.04   &       0.34$\pm$0.05    &      0.46$\pm$0.04    \\
   \\
                       
Ni$_{80}$Fe$_{20}$ (5 nm)    &        0.00$\pm$0.00   &      0.00$\pm$0.00&      8.83$\pm$0.54 &       0.00$\pm$0.00  &      0.00$\pm$0.00   &      5.31$\pm$0.48   &       0.35$\pm$0.02    &       0.32$\pm$0.05   &      0.50$\pm$0.06   \\
       \\
                    \\ \\ \hline 
\end{tabular}}
\end{table*}
The Raman spectra of CVD grown ML-$\mathrm{MoS_{2}}$ on a Si/SiO$_{2}$ substrate before and after the deposition of Pt(1 nm)/Py(5 nm) and Pt(2 nm)/Py(5 nm) bilayer samples are shown in Fig.~\ref{fig1}a. These spectra shows that the quality of ML-$\mathrm{MoS_{2}}$ layer remains unchanged even after the deposition of IL Pt of different thicknesses and Py(5 nm) layers, as we do not see any change in the linewidth of the Raman peaks. We do not observe any additional peaks after the deposition of Pt and Py, which is often observed when the disorder is introduced into the $\mathrm{MoS_{2}}$ layer~\cite{Mignuzzi}. Fig.~\ref{fig1}a shows the Raman spectrum that highlights two well known signatures of $\mathrm{MoS_{2}}$: $\mathrm{E_{2g}}$ and $\mathrm{A_{1g}}$ are observed at $\approx$ 384.54 $\mathrm{cm}^{-1}$ and $\approx$  405.31 $\mathrm{cm}^{-1}$, respectively. The difference in wave number or Raman shift ($\mathrm{\delta}$) $\approx$  20.77 $\mathrm{cm}^{-1}$ between two peaks $\mathrm{E_{2g}}$ and $\mathrm{A_{1g}}$ confirms the monolayer nature of $\mathrm{MoS_{2}}$~\cite{Zhang}. Monolayer of $\mathrm{MoS_{2}}$, belonging to the $\mathrm{P6m2}$ space group, comprises a Mo layer sandwiched between two S layers, forming a three atom unit cell. The $\mathrm{A_{1g}}$ mode involves out of plane vibrations of S atoms in opposite directions, while the $\mathrm{E_{2g}}$ mode corresponds to in plane counter vibrations of S atoms relative to Mo~\cite{Wieting,Verble}.

To check the quality of the thin film samples, the deposited stacks are subjected to X-ray reflectivity (XRR) measurements for accurate estimation of the thickness and interface roughness using a smartlab Rigaku X-ray diffractometer. The simulation of the recorded spectra is done by using X-ray reflectivity software (segmented V.1.2) considering a stack of Si/SiO$_{2}$/ML-MoS$_{2}$/Pt(0,1,2,5 nm)/Py(5 nm). The XRR spectra of the thin film stacks are recorded to estimate their exact thickness, density, and interface roughness. The discernible Kiessig fringes throughout the complete measurement range demonstrate that the thin films and their interfaces have excellent quality (Fig.~\ref{fig1}b). The low interface roughness of the films, in addition to the fact that the thickness of each layer closely resembles the value obtained from deposition rate calibration, indicates that there is minimal intermixing of FM and NM during the fabrication of thin film stacks. The estimated values for each layer's thickness, density, and interface roughness are displayed in Table~\ref{table1}. The thickness of ML-$\mathrm{MoS_2}$ is observed to be $\approx 0.90 \pm 0.09$ nm, aligning well with the reported values~\cite{NKGupta}. The simulated density of each layer is also found to be close to their bulk values. A rough FM/NM interface affects spin transport channels since it influences both spin impedance matching and the interfacial SOC~\cite{Jin}. The saturation magnetization of all the thin film stacks is measured by a SQUID-based magnetometer (MPMS 3, Quantum Design). The typical in-plane (open black squares) and out of plane (open red circles) magnetic hysteresis loops for Si/SiO$_{2}$/ML-MoS$_{2}$/Py(5 nm) samples measured at room temperature are shown in Fig.~\ref{fig1}c. The saturation magnetization ($\mathrm{M_{s}}$) value determined from the magnetic hysteresis measurement for the Si/SiO$_{2}$/ML-MoS$_{2}$/Py(5 nm) sample is found to be $\approx$ 800 mT.

\section{Magnetization dynamics and spin pumping}

In order to investigate the magnetization dynamics and spin pumping, a lock in based broadband ferromagnetic resonance (FMR) set up (NanOsc) is utilized. The thin film samples of size $4 \times 4$ mm$^2$ are placed on a 200 $\mu$m wide coplaner waveguide (CPW) in a flip chip manner and the FMR measurements are carried out in the broad frequency range (3–15 GHz) as a function of in plane applied effective DC magnetic fields in a direction perpendicular to the radio frequency field (h$_{\mathrm{rf}}$) at room temperature~\cite{Tahir1,Tahir2}. To estimate the in plane (IP) and out of plane (OOP) anisotropies we perform in plane azimuthal angle dependent FMR measurements for all samples.
\begin{figure*}
\includegraphics[width=0.99\linewidth]{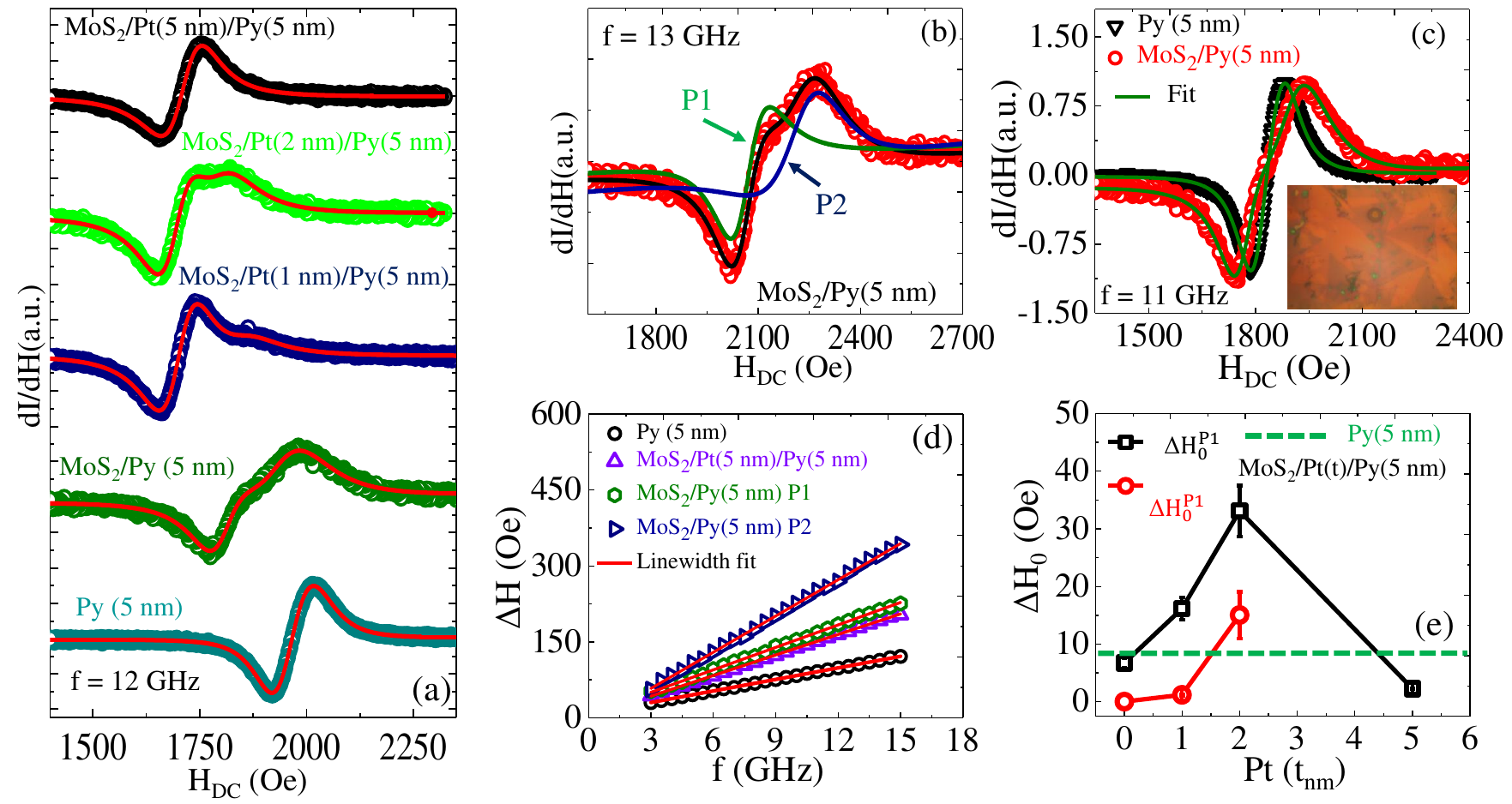}
\caption{(a) FMR spectra of Py(5 nm), ML-MoS$_{2}$/Py(5 nm), and ML-MoS$_{2}$/Pt(1,2,5 nm)/Py(5 nm) at 12 GHz, with experimental data (symbols) and Lorentzian fits (red lines, Eq.~\ref{eq1}). (b) Two peak FMR spectra for ML-MoS$_{2}$/Py(5 nm) at 13 GHz, fitted using a double Lorentzian function. (c) FMR spectra of Si/SiO$_{2}$/Py(5 nm) and Si/SiO$_{2}$/ML-MoS$_{2}$/Py(5 nm) at 11 GHz, with data points (symbols) and Lorentzian fits (solid lines). Inset: optical image of ML-MoS$_{2}$ after Py deposition. (d) Linewidth ($\Delta \mathrm{H}$) vs. frequency ($\mathrm{f}$) for Si/SiO$_{2}$/Py(5 nm), Si/SiO$_{2}$/ML-MoS$_{2}$/Pt(5 nm)/Py(5 nm), and both peaks of Si/SiO$_{2}$/ML-MoS$_{2}$/Py(5 nm), with fits using Eq.~\ref{eq3}. (e) Inhomogeneous linewidth broadening ($\Delta \mathrm{H_{0}}$) vs. IL Pt thickness, with first and second peaks shown as black and red data points, respectively.}
\label{fig2}
\end{figure*}

The FMR measurements are performed in applied effective magnetic fields $\mathrm{H}_\mathrm{eff}$ well above the in plane anisotropy fields, facilitating the magnetization $\mathrm{M}$ of the ferromagnetic layer to be considered parallel to the applied effective magnetic field. As shown in Fig.~\ref{fig1}d, when the applied microwave magnetic field acts on the FM layer to cause ferromagnetic resonance, the magnetization M around the effective magnetic fields $\mathrm{H}_\mathrm{eff}$ is transmitted to the adjacent NM layer. The spin polarization of electrons in the NM layer is caused by the spin transfer torque due to the compensation of spin angular momentum. This results in the formation of a pure spin current along the z-direction and a spin polarization vector along the direction of the applied effective magnetic field $\mathrm{H}_\mathrm{eff}$. The FMR spectrum develops two peaks with the introduction of MoS$_2$ ML which merges into a single peak when the Pt IL thickness is increased to 5 nm (Fig.~\ref{fig2}a). The FMR spectra are fitted by the derivative Lorentzian function given by Eq.~\ref{eq1}) having symmetric and asymmetric amplitudes~\cite{Woltersdorf}. The deconvoluted FMR spectra for ML-MoS$_{2}$/Py(5 nm) interface without the Pt IL are shown separately in Fig.~\ref{fig2}b.
\begin{widetext}
\begin{equation}
\frac{d \mathrm{I}_{\mathrm{FMR}}}{d \mathrm{H}} = \sum_{i=1,2}4\mathrm{A_i}\frac{\Delta \mathrm{H_i}( \mathrm{H}- \mathrm{H}_{\mathrm{resi}})}{(4( \mathrm{H}- \mathrm{H}_{\mathrm{resi}})^2+
(\Delta  \mathrm{H_i})^2)^2}-\mathrm{S_i}\frac{(\Delta  \mathrm{H_i})^2-4(\mathrm{H}-\mathrm{H}_{\mathrm{resi}})^2}{(4( \mathrm{H}- \mathrm{H}_{\mathrm{resi}})^2+(\Delta  \mathrm{H_i})^2)^2}\label{eq1}
\end{equation}
\end{widetext}

Here $\mathrm{H}$, $\Delta\mathrm{H}$, and $\mathrm{H}_{\mathrm{res}}$ are the in plane applied DC magnetic field, FMR linewidth, and resonance field of microwave absorption, respectively. The amplitudes $\mathrm{S}$ and $\mathrm{A}$ of the FMR signal are associated with symmetric and 
antisymmetric coefficients, respectively~\cite{Woltersdorf}. What is the origin of the two peak feature in the FMR spectra with MoS$_2$ interface? We find that the single crystalline monolayer (ML) of MoS$_{2}$ is not continuous; instead, it consists of monolayer islands of triangle shape with lateral dimensions up to several tens of $\mu$m, as can be observed from the optical microscope image shown in the inset of Fig.~\ref{fig2}c. We attribute the origin of the first peak (P1) of FMR spectra to the Py layer directly in contact with the Si/SiO$_{2}$ substrate i.e. the portions uncovered by the ML-MoS$_{2}$ islands, while we ascribe the second peak (P2) to the Py layer deposited on top of the ML-MoS$_{2}$ islands. The two interfaces are responsible for the distinct two peak features observed in the FMR spectra of the ML-MoS$_{2}$/Pt(0,1,2 nm)/Py(5 nm) samples. The two peak characteristics completely disappear with the Pt(5 nm) interlayer, as the uncovered portions between the ML-MoS$_{2}$ islands are filled by the Pt, resulting in the FM film being deposited on the continuous Pt(5 nm) interface. As shown in Fig.~\ref{fig2}a, the ML-MoS$_{2}$/Pt(5 nm)/Py(5 nm) sample exhibits a single FMR spectrum, similar to that of the Py(5 nm) sample. To further elucidate the complex line shape observed at the ML-MoS$_{2}$/Pt(0,1,2 nm)/Py(5 nm) interfaces, we conduct a systematic study across a broad frequency range of 3 to 15 GHz. Even at high frequencies, the two peaks observed in the FMR spectra are distinguishable. For each FMR spectra observed for ML-MoS$_{2}$/Pt(0,1,2 nm)/Py(5 nm) interfaces, two distinct $\Delta \mathrm{H}$ and $\mathrm{H}_{\mathrm{res}}$ have been extracted from the Lorentzian fittings. Figure~\ref{fig2}c shows the FMR spectra of Py(5 nm) and ML-MoS$_{2}$/Py(5 nm) samples recorded at 11 GHz. The linewidth $\Delta \mathrm{H}$ increases from 148.77 Oe for Si/SiO$_{2}$/Py(5 nm) to 162.36 Oe for the first peak and further to 264.56 Oe at the Si/SiO$_{2}$/ML-MoS$_{2}$/Py(5 nm) interface. The observed enhancement in the $\Delta \mathrm{H}$ is attributed to the transfer of spin angular momentum from the Py(5 nm) layer to the ML-MoS$_{2}$ layer.

The surface and interface imperfections, as well as defects in the FM thin films, can contribute to an increased linewidth in the FMR spectra. The total FMR linewidth comprises both intrinsic and extrinsic contributions and can be expressed as:

\begin{figure*}
\includegraphics[width=0.99\linewidth]{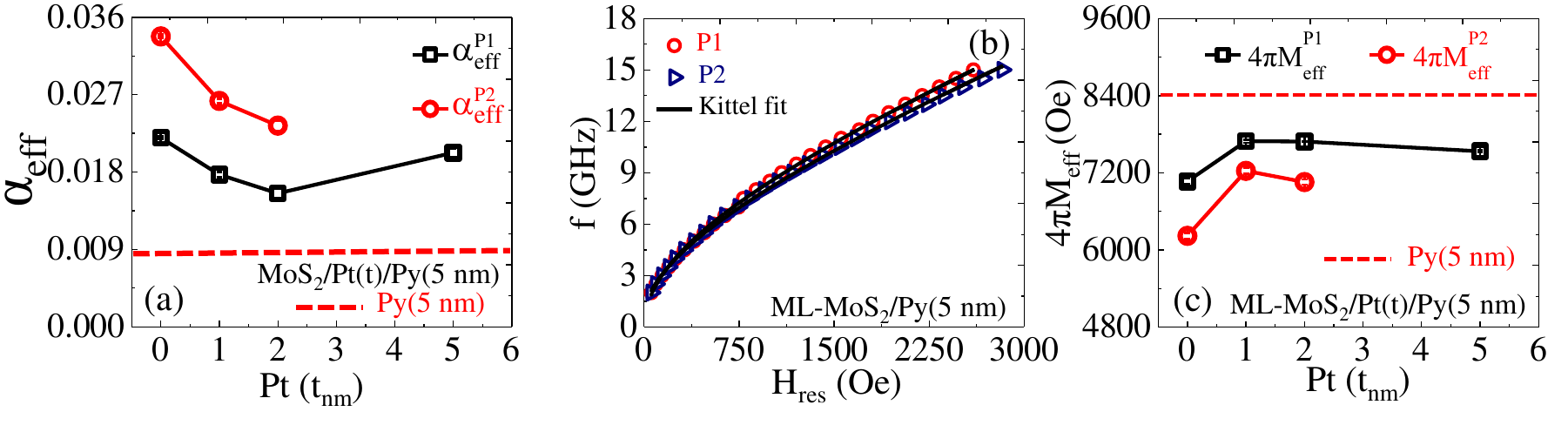}
\caption{(a) The effective Gilbert damping parameter ($\alpha_{\mathrm{eff}}$) as a function of IL Pt thickness, where black and red data points represent the first and second FMR peaks, respectively. The red dotted line indicates $\alpha_{\mathrm{eff}}$ for the reference Py (5 nm) sample. (b) The resonance field ($\mathrm{H}_{\mathrm{res}}$) as a function of frequency ($\mathrm{f}$) for ML-MoS$_{2}$/Py (5 nm), with black solid lines representing Kittel fits (Eq.~\ref{eq5}). (c) The effective magnetization ($4\pi\mathrm{M}_\mathrm{eff}$) plotted as a function of IL Pt thickness.}  
\label{fig3}
\end{figure*}
\begin{eqnarray}
\Delta \mathrm{H}= \Delta\mathrm{H}_\mathrm{intrinsic}+\Delta\mathrm{H}_\mathrm{extrinsic}\label{eq2}
\end{eqnarray}

In the case of conduction electrons, the intrinsic linewidth arises from spin orbital coupling, as represented by the Gilbert damping contributions~\cite{Hickey}. The extrinsic linewidth is a frequency independent feature associated with the magnetic inhomogeneity of a thin film, commonly referred to as inhomogeneous linewidth broadening. The extrinsic contributions are determined by the internal magnetic field, or the local fluctuation in magnetization~\cite{Kittel, Gilbert}. 
Fig.~\ref{fig2}d shows the variation of $\Delta \mathrm{H}$ as a function of $\mathrm{f}$ for Py (5 nm), as well as for the first and second peaks of the ML-MoS$_{2}$/Py(5 nm) interface and the ML-MoS$_{2}$/Pt(5 nm)/Py(5 nm) interface. The solid line fit follows the damping equation given by:
\begin{eqnarray}
\Delta \mathrm{H}= \frac{4\pi \alpha_{\mathrm{eff}}}{\gamma} \mathrm{f}+\Delta\mathrm{H}_0\label{eq3}
\end{eqnarray}
The first term accounts for the frequency dependent magnetization relaxation mechanism, while the second term, $\Delta\mathrm{H}_0$, represents the frequency independent contribution associated with inhomogeneous linewidth broadening~\cite{Kittel, Gilbert}. Figure~\ref{fig2}d shows that the linear behavior of $\Delta \mathrm{H}$ versus $\mathrm{f}$ indicates an intrinsic origin of the damping parameter observed in our samples. The extracted values of $\Delta\mathrm{H}_0$, derived from the first and second peaks of the FMR spectra, reflect the two different interfaces formed in each sample, as discussed earlier. The inhomogeneous linewidth broadening $\Delta \mathrm{H_{0}}$ values for the ML-MoS$_{2}$/Pt(0,1,2,5 nm)/Py(5 nm) heterostructures, along with the reference Py(5 nm) sample, are shown in Fig.~\ref{fig2}e. The observed variations in $\Delta \mathrm{H}_{0}$ across different MoS$_{2}$/Pt(0,1,2,5 nm)/Py(5 nm) interfaces highlight distinct interfacial characteristics. The consistently low $\Delta \mathrm{H}_{0}$ values across most samples indicate the sustained high structural and magnetic quality of each sample. However, the slightly higher $\Delta \mathrm{H}_{0}$ for MoS$_{2}$/Pt(2 nm)/Py(5 nm) suggests that the 2 nm Pt interlayer plays a more significant role in modifying the interface. These variations provide strong evidence of interface specific differences, likely originating from variations in interfacial coupling and spin transport mechanisms.

The FMR linewidth for intrinsic damping is expected to depend linearly on the microwave frequency, according to the Landau Lifshitz Gilbert equation~\cite{Kittel, Gilbert}. In the present scenario, the values of $\alpha_{\mathrm{eff}}$ shown in Fig.~\ref{fig3}a are determined by utilizing the slope of the linear frequency dependence of linewidth consist of the following contributions:
\begin{eqnarray}
 \alpha_\mathrm{eff} = \alpha_\mathrm{Py}+\alpha_\mathrm{SP}\label{eq4}
\end{eqnarray}
Here $\alpha_\mathrm{Py}$ is the Gilbert damping parameter of the reference Py layer, which arises due to the energy transfer to the lattice within the bulk, resulting in the relaxation of spin angular momentum within the FM lattice~\cite{Hickey}. The damping constant, $\alpha_\mathrm{SP}$, arises from the loss of spin angular momentum due to the outflow of spin current from the FM layer into the NM layer, as well as from spin flip processes at the interface caused by interfacial spin orbit coupling. The extracted values of the effective Gilbert damping parameter ($\alpha_\mathrm{eff}$) for the first and second peaks of the ML-MoS$_{2}$/Pt(0,1,2 nm)/Py and ML-MoS$_{2}$/Pt(5 nm)/Py interfaces are found to be significantly higher than that of the reference Py layer. The observed values of $\alpha_{\mathrm{eff}}$ as a function of interlayer Pt thickness are plotted in Fig.~\ref{fig3}a. This enhancement in $\alpha_\mathrm{eff}$ is attributed to the transfer of spin angular momentum from the Py layer to the adjacent high SOC Pt(0,1,2,5 nm)/ML-MoS$_{2}$ interfaces~\cite{Conca}.

Fig.~\ref{fig3}b shows the extracted resonance field $\mathrm{H}_{\mathrm{res}}$ values for each microwave absorption frequency $\mathrm{f}$ corresponding to the first and second peak of FMR spectra observed in ML-MoS$_{2}$/Py interface. The effective magnetization $4\pi \mathrm{M}_{\mathrm{eff}}$ for the ref. Py(5 nm) sample and Si/SiO$_{2}$/ML-MoS$_{2}$/Pt(0,1,2,5 nm)/Py(5 nm) interfaces are calculated as a function of IL Pt thickness using Kittel’s equation (Eq.~\ref{eq5})~\cite{Kittel}. 
\begin{equation}
\mathrm{f} = \frac{\gamma}{2\pi}\left[\left( \mathrm{H}_{\mathrm{res}}+ \mathrm{H}_{\mathrm{k}}\right) \left( \mathrm{H}_{\mathrm{res}}+ \mathrm{H}_{\mathrm{k}}+ 4 \pi \mathrm{M}_{\mathrm{eff}}\right)\right]^{1/2}\label{eq5}
\end{equation}
Here $\gamma$ = $\frac{g\mu_{B}}{\hbar}$ = $1.88 \times 10^{2}$ GHz/T is the gyromagnetic ratio, $\mu_{B}$ is the Bohr magneton, $g$ is the lande's spectroscopic splitting factor, and $\hbar$ is the reduced Planck's constant, $\mathrm{H}_{\mathrm{k}}$ is the in plane anisotropy field of the FM layer.
Figure~\ref{fig3}c shows the variation of $4 \pi \mathrm{M}_{\mathrm{eff}}$
  with Pt interlayer thickness. The Py (5 nm) sample exhibits a higher $4 \pi \mathrm{M}_{\mathrm{eff}}$
compared to the values obtained from the first and second FMR peaks in ML-MoS$_{2}$/Pt(0,1,2 nm)/Py(5 nm) interfaces and ML-MoS$_{2}$/Pt(5 nm)/Py(5 nm) interface. The effective magnetization, $4\pi\mathrm{M}_{\mathrm{eff}}$, is expressed as:
\begin{equation}
4\pi\mathrm{M}_{\mathrm{eff}} = 4\pi\mathrm{M}_{\mathrm{s}}- \mathrm{K}_{\mathrm{s}}/\mathrm{M}_{\mathrm{s}}\mathrm{t}_{\mathrm{Py}}\label{eq6}
\end{equation}
where $\mathrm{K}_{\mathrm{s}}$ is the surface/interface anisotropy constant, $\mathrm{t}_{\mathrm{Py}}$ is the thickness of the Py layer, and  $4\pi\mathrm{M}_{\mathrm{s}}$ is the saturation magnetization. The reduction of $4\pi \mathrm{M}_{\mathrm{eff}}$ in the ML-MoS$_{2}$/Pt(1,2,5 nm)/Py(5 nm) system is explained by enhanced interfacial SOC and magnetic interactions in the multilayer stack. The $4\pi \mathrm{M}_{\mathrm{eff}}$ values reflect both the intrinsic properties of the FM layer and the extrinsic influences from the adjacent NM layers. Introducing Pt, with high SOC, at the interface modifies both interfacial and in plane anisotropy fields, affecting the demagnetizing field and resulting in reducing  $4 \pi \mathrm{M}_{\mathrm{eff}}$ values for ML-MoS$_{2}$/Pt(1,2,5 nm)/Py(5 nm) interfaces, compared to the ref. Py(5 nm) sample. The decrease in $4\pi\mathrm{M}_{\mathrm{eff}}$ for ML-MoS$_{2}$/Py(5 nm) interface arises from increased interfacial spin orbit coupling, which is induced by d-d hybridization~\cite{Wu,Jamilpanah}. This is consistent with prior reports on 2D material/FM interfaces, where similar mechanisms were found to affect magnetic properties~\cite{Wu,Jamilpanah}. Two slightly different $4 \pi \mathrm{M}_{\mathrm{eff}}$ values are observed due to the two different interfaces of ML-MoS$_{2}$/Pt(0,1,2 nm)/Py(5 nm) samples as shown in Fig.~\ref{fig3}c.
\begin{figure*}
\includegraphics[width=\linewidth]{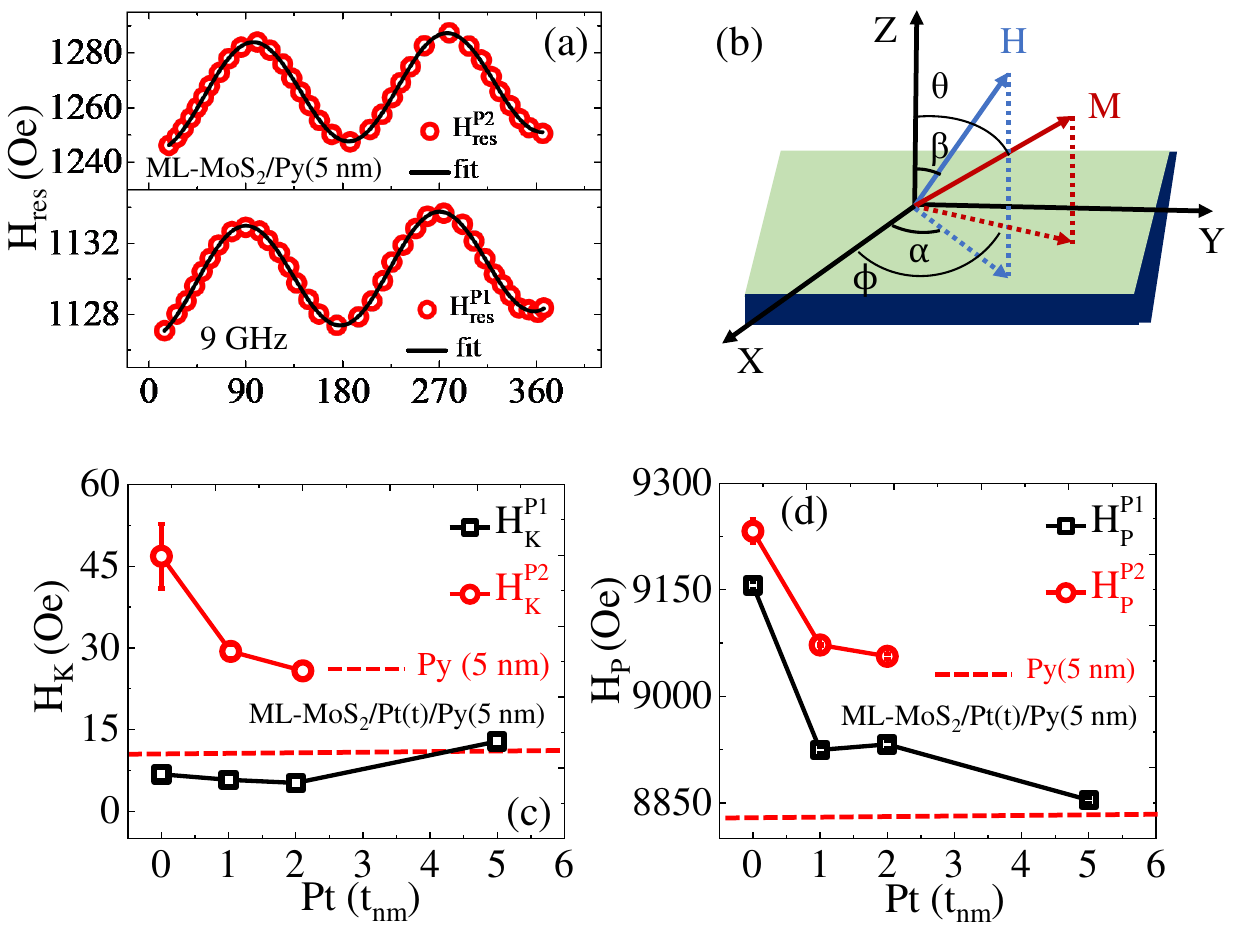}
\caption{(a) Azimuthal angle ($\phi$) dependence of $\mathrm{H}_{\mathrm{res}}$ at $\mathrm{f} = 9$ GHz, exhibiting a sinusoidal trend, indicative of both in-plane (IP) and out-of-plane (OOP) anisotropies at both resonance peaks. The solid line represents a fit using Eq.~\ref{eq10} to extract $\mathrm{H}_{\mathrm{k}}$ and $\mathrm{H}_{\mathrm{p}}$. (b) Schematic illustrating magnetization ($\mathrm{M}$) and external field ($\mathrm{H}$) in an in-plane magnetized film, where ($\alpha$, $\phi$) denote azimuthal angles and ($\beta$, $\theta$) represent polar angles relative to the film plane. (c) and (d) Dependence of in-plane ($\mathrm{H}_{\mathrm{k}}$) and out-of-plane ($\mathrm{H}_{\mathrm{p}}$) anisotropies on Pt interlayer thickness ($\mathrm{t}_{\mathrm{Pt}}$), extracted from the first and second FMR peaks of the Py layer.}  
\label{fig4}
\end{figure*}

To estimate the interlayer thickness dependent magnetic anisotropies, we performed in plane azimuthal angle dependent FMR measurements on ref. Py and ML-MoS$_{2}$/Pt(0, 1, 2,5 nm)/Py samples at 9 GHz. The correlation between the azimuthal angle of the magnetization vector in an in plane magnetized film and the resonant magnetic field of FMR is utilized to determine the magnetic anisotropic fields associated with perpendicular magnetic anisotropy (PMA) and in plane magnetic anisotropy (IMA). The in plane angle dependent FMR measurement is a reliable technique for determining magnetic anisotropies in FM thin films~\cite{Deka}. This is especially true when the magnetic anisotropy field is much smaller than the demagnetizing field. Moreover, to avoid nonlinear magnetization dynamics that can alter the demagnetizing field during FMR measurements, we must use low input power. This approach allows for the precise determination of anisotropy fields down to a few Oersteds~\cite{Medwal}. We assume that the direction of the external magnetic field ($\mathrm{H}_{\mathrm{DC}}$) and the magnetization $\mathrm{M}$ is specified by (${\phi}$ , ${\theta}$) and (${\alpha}$, ${\beta}$), respectively, where (${\alpha}$, ${\phi}$) are the azimuthal angles and (${\beta}$, ${\theta}$) are the polar angles concerning the film plane as shown in Fig.~\ref{fig4}b. When the Zeeman energy, demagnetization energy, and magnetic anisotropy energy are taken into account, the magnetic free energy $\mathrm{F}$ is described as follows~\cite{Gallardo}:
\begin{widetext}
\begin{equation}
 \mathrm{F} = \frac{{\mathrm{M}_{\mathrm{s}}}}{2}[-2\mathrm{H}_{\mathrm{ex}}(\cos{\theta}\cos{\beta}\cos{(\alpha-\phi)+
 \sin{\theta}\sin{\beta}})+\\(\mathrm{M}_{\mathrm{s}}-\mathrm{H}_{\mathrm{p}})\sin^2{\theta}-\mathrm{H}_{\mathrm{k}}\cos^2{\theta}\cos^2{\phi}]\label{eq7}
\end{equation}
\end{widetext}
Where $\mathrm{H}_{\mathrm{k}}$ represents the in plane anisotropy field directed along $\phi = \theta = 0^\circ$, and $\mathrm{H}_{\mathrm{P}}$ represents the perpendicular magnetic anisotropy field along $\theta = 90^\circ$.

Next, we substitute $\mathrm{F}$ into the Smit-Beljers relation~\cite{Smit}, given by:
\begin{equation}
{\left(\frac{\mathrm{f}}{\gamma}\right)^2}=\frac{1}{\left(\mathrm{M}_{\mathrm{s}}\cos\theta\right)^{2}}\left[{\frac{\partial^2\mathrm{F}}{\partial\theta^2}}{\frac{\partial^2\mathrm{F}}
{\partial\phi^2}}-\left(\frac{\partial^2\mathrm{F}}{\partial\theta \partial\phi}\right)^2\right]
\label{eq8}
\end{equation}
\begin{figure*}
\includegraphics[width=\linewidth]{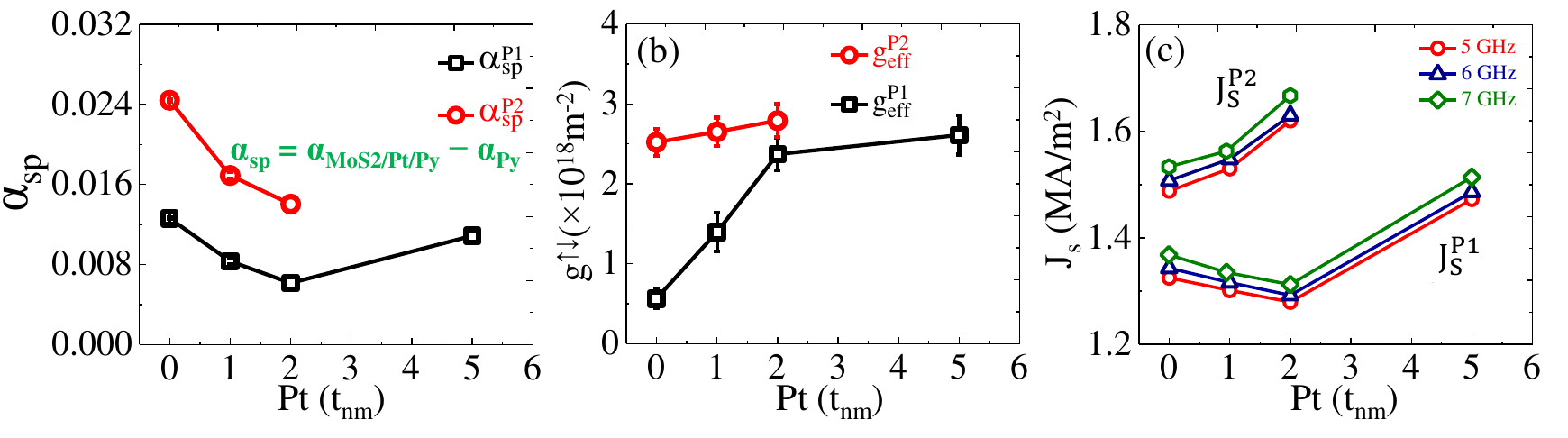}
\caption{(a) Spin pumping induced Gilbert damping factor ($\alpha_{\mathrm{sp}}$) as a function of Pt interlayer thickness ($\mathrm{t}_{\mathrm{Pt}}$). (b) Interfacial spin-mixing conductance ($\mathrm{g}^{\uparrow\downarrow}$) extracted from the first and second FMR peaks of ML-MoS$_{2}$/Pt(0,1,2,5 nm)/Py(5 nm) as a function of $\mathrm{t}_{\mathrm{Pt}}$, evaluated using Eq.~\ref{eq11}. (c) Effective spin-current density injected into Pt(0,1,2,5 nm)/ML-MoS$_{2}$ vs. $\mathrm{t}_{\mathrm{Pt}}$, calculated at $\mathrm{f} = 5,6,7$ GHz using Eq.~\ref{eq12}.}  
\label{fig5}
\end{figure*}
In this study, in plane FMR measurements are carried out while $\mathrm{M}$ remains parallel to the $\mathrm{H}_{\mathrm{ex}}$ to determine the $\mathrm{H}_{\mathrm{P}}$ and $\mathrm{H}_{\mathrm{k}}$. Using $\theta = \beta = 0^{\circ}$ and $\phi = \alpha$, we obtain the following relation:
\begin{widetext}
\begin{equation}
{\left(\frac{\mathrm{f}}{\gamma}\right)^2} = \frac{1}{\mathrm{M}_{\mathrm{s}}^{2}}\left[{\mu_{0}{\mathrm{M}_{\mathrm{s}}}}\mathrm{H}_{\mathrm{ex}}+{\mu_{0}{\mathrm{M}_{\mathrm{s}}}}\left(\mathrm{M}_{\mathrm{s}}-\mathrm{H}_{\mathrm{p}}\right)+{\mu_{0}{\mathrm{M}_{\mathrm{s}}}}\mathrm{H}_{\mathrm{k}}\cos^2{\phi}\right]
\left[{\mu_{0}{\mathrm{M}_{\mathrm{s}}}}\mathrm{H}_{\mathrm{ex}}+{\mu_{0}{\mathrm{M}_{\mathrm{s}}}}\mathrm{H}_{\mathrm{k}}\cos{2\phi}\right]
\label{eq9}
\end{equation}
\end{widetext}
Replacing $\mathrm{H}_{\mathrm{ex}}$ with $\mathrm{H}_{\mathrm{res}}$ in the above equation, we obtain the relation between $\mathrm{H}_{\mathrm{res}}$ and ${\phi}$ as follows.

\begin{widetext}
\begin{equation}
 \mathrm{H}_{\mathrm{res}} = -\mathrm{H}_{\mathrm{k}}+\frac{3}{2}\mathrm{H}_{\mathrm{k}}\sin^2{(\phi+\Delta)}-\frac{\mathrm{M}_{\mathrm{s}}-\mathrm{H}_{\mathrm{p}}}{2}+\frac{1}{2}\left[\mathrm{H}^2_{\mathrm{k}}\sin^4{(\phi+\Delta)} +(\mathrm{M}_{\mathrm{s}}-\mathrm{H}_{\mathrm{p}})^2 +2(\mathrm{M}_{\mathrm{s}}-\mathrm{H}_{\mathrm{p}}) \mathrm{H}_{\mathrm{k}}\sin^2{(\phi+\Delta)}+\left(\frac{2\mathrm{f}}{\mu_{0}\gamma}\right)^2\right]^{\frac{1}{2}}\label{eq10}
\end{equation}
\end{widetext}

Here $\Delta$ accounts for the offset in the magnitude of the lowest or the highest resonance field. In order to estimate the in plane (IP) anisotropy field $\mathrm{H}_{\mathrm{k}}$ and the out of plane (OOP) anisotropy field  $\mathrm{H}_{\mathrm{P}}$, in plane FMR spectra for the ML-MoS$_{2}$/Pt(0,1,2,5 nm)/Py(5 nm) interfaces are recorded at different azimuthal angles $\phi$ that ranges from $0^{\circ}$ to $360^{\circ}$. The resonance field $\mathrm{H}_{\mathrm{res}}$ observed due to the first and second peak as a function of in plane angle $\phi$ for $\mathrm{f}$ = 9 GHz are shown in the lower and upper panel of Fig.~\ref{fig4}a. The data is fitted with the expression in Eq.~\ref{eq10}, using the saturation magnetization ($\mathrm{M}_{\mathrm{s}}$ = 800 mT) and frequency ($\mathrm{f}$ = 9 GHz) as fixed parameters. This fitting allows us to estimate the values of $\mathrm{H}_{\mathrm{k}}$ and $\mathrm{H}_{\mathrm{p}}$  anisotropies observed for first and second peak of Py layer. For $\mathrm{f} = 9$ GHz, we measure the $\mathrm{H}_{\mathrm{k}}$ and $\mathrm{H}_{\mathrm{P}}$ due to first and second peak as a function of interlayer Pt thickness as shown in Fig.~\ref{fig4}c and Fig.~\ref{fig4}d, respectively. Fig.~\ref{fig4}c presents the dependence of the in plane anisotropy field ($\mathrm{H}_{\mathrm{k}}$) on Pt interlayer thickness ($\mathrm{t}_{\mathrm{Pt}}$) for ML-MoS$_{2}$/Pt(0,1,2,5 nm)/Py(5 nm) heterostructures. In the reference Py(5 nm) sample, $\mathrm{H}_{\mathrm{k}}$ remains relatively low, consistent with the weak magnetocrystalline anisotropy and low coercivity of Py. In contrast, ML-MoS$_{2}$/Pt(0,1,2,5 nm)/Py(5 nm) systems exhibit a notable enhancement in $\mathrm{H}_{\mathrm{k}}$, indicating modifications at the interface due to Py deposition on ML-MoS$_{2}$ islands and the ML-MoS$_{2}$/Pt surface. This increase strongly suggests the influence of interfacial exchange interactions and spin orbit coupling (SOC) effects at the ML-MoS$_{2}$/Pt interface~\cite{Husain,Thiruvengadam}. With increasing Pt interlayer thickness, $\mathrm{H}_{\mathrm{k}}$ enhances while ($\mathrm{H}_{\mathrm{P}}$) reduces, indicating the modulation of interfacial spin interactions and anisotropic energy contributions.

The large increase in $\alpha _{\mathrm{eff}}$ is observed for the ML-MoS$_{2}$/Pt(0,1,2 nm)/Py(5 nm) interface compared to the reference  Py(5 nm) sample shown in Fig.~\ref{fig3}a. This enhancement in the Gilbert damping is attributed to transfer of spin angular momentum from FM to NM layer and is estimated by $\alpha _{\mathrm{sp}}$ = $\Delta\alpha = \alpha_{\mathrm{MoS_{2}/Pt(t)/Py}} - \alpha_{\mathrm{Py}}$ shown in Fig.~\ref{fig5}a. The expression is utilised to compute the interfacial spin mixing conductance ($\mathrm{g}^{\uparrow\downarrow}$) which is given by the following expression:~\cite{Tser2, Tser3}.
\begin{eqnarray}
\mathrm{g}^{\uparrow\downarrow} =\frac{4\pi \mathrm{M}_{s}\mathrm{t}_{\mathrm{Py}}}{\mathrm{g}\mu_{B}}(\alpha_{\mathrm{MoS_{2}/Pt(t)/Py}}-\alpha_{\mathrm{Py}})\label{eq11}
\end{eqnarray}
Here, $\mu_{B}$ is the Bohr magneton and $4 \pi\mathrm{M}_{\mathrm{s}}$ is the saturation magnetization. The extracted values of ($\mathrm{g}^{\uparrow\downarrow}$) at the ML-MoS$_{2}$/Py(5 nm) interface is found to be increasing after the interface modification through high SOC material Pt(1,2,5 nm) as an interlayer. The values of ($\mathrm{g}^{\uparrow\downarrow}$) for ML-MoS$_{2}$/Py interface is found to be $0.52 \times 10^{18}$ m$^{-2}$ which is further increases to $2.52 \times 10^{18}$ m$^{-2}$ for the MoS$_{2}$/Py(5 nm) interface with respect to the Pt (1,2,5 nm) interlayer, as shown in Fig.~\ref{fig5}b, and remains consistent with the reported values for the MoS$_{2}$/FM interface.

The reported values of spin-mixing conductance ($\mathrm{g}^{\uparrow\downarrow}$) in MoS$_{2}$/FM systems with different FMs are $1.54 \times 10^{19}$ m$^{-2}$ for MoS$_{2}$/Co~\cite{Cheng}, $ 0.72 \times 10^{19}$ m$^{-2}$ for the MoS$_{2}$/Ni$_{81}$Fe$_{19}$~\cite{Bangar} interface, $2.21 \times 10^{19}$ m$^{-2}$ for the MoS$_{2}$/Co$_{60}$Fe$_{20}$B$_{20}$~\cite{Gupta} interface, $1.49 \times 10^{19}$ m$^{-2}$ for the MoS$_{2}$/Co$_{2}$FeAl interface~\cite{Husain}, and $0.4 \times 10^{19}$ m$^{-2}$ for the MoS$_{2}$/Co$_{2}$FeSi interface~\cite{Sharma1}. It is important to note that the spin diffusion length in MoS$_{2}$ is very small, approximately $0.64 \pm 0.25$ nm \cite{Husain}. Despite this small spin diffusion length, the spin mixing conductance in MoS$_{2}$ is comparable to that observed in heavy metal (HM) systems. Reported values of spin mixing conductance for various ferromagnet FM/HM stacks are as follows: $1.16 \times 10^{19}$ m$^{-2}$ for Mo/$\mathrm{Co_2FeAl}$ \cite{Chaudhary}, $1 \times 10^{19}$ to $4 \times 10^{19}$ m$^{-2}$ for Co/Pt \cite{Azzawi,Rojas}, $9.7 \times 10^{19}$ m$^{-2}$ for YIG/Pt \cite{Haertinger}, and $10 \times 10^{19}$ m$^{-2}$ for Fe/Pd \cite{Kumar}. Therefore, the utilization of one ML-MoS$_{2}$ in a FM/NM heterostructure provides outstanding properties that could be potentially useful for spin dynamics applications.

The enhancement of the Gilbert damping and the interfacial spin mixing conductance observed in the ML-MoS$_{2}$/Pt(0,1,2,5 nm)/Py(5 nm) stacks could be attributed to: 1) the spin current injected in the ML-MoS$_{2}$ by the spin pumping mechanism at the ML-MoS$_{2}$/Pt(0,1,2 nm)/Py(5 nm) interface, which leads to spin accumulation on the ML-MoS$_{2}$. 2) The dissipation of spin current at the ML-MoS$_{2}$/Py and ML-MoS$_{2}$/Pt(0,1,2,5 nm)/Py interfaces through spin-flip scattering acts as an additional channel for spin relaxation, leading to enhanced damping $\alpha_{\mathrm{sp}}$~\cite{Ando1}. 
The diffusive flow of spins in ML-MoS$_{2}$/Pt(0,1,2,5 nm)/Py can be described by spin current density $\mathrm{J}_{\mathrm{s}}$, evaluated using the following expression~\cite{Tser1,Tser2}:
{\begin{widetext}
\begin{equation}
\mathrm{J}_{\mathrm{s}} \approx \left(\frac{\mathrm{g}^{\uparrow\downarrow}\hbar}{8\pi}\right)\left(\frac{
\mathrm{h}_{\mathrm{rf}}\gamma}{\alpha}\right)^2 \left[\frac{4\pi \mathrm{M}_{\mathrm{s}}\gamma+\sqrt{(4\pi\mathrm{M}_{\mathrm{s}}\gamma)^2+16(\pi \mathrm{f})^2}}{(4\pi\mathrm{M}_{\mathrm{s}}\gamma)^2+16(\pi \mathrm{f})^2}\right]\left(\frac{2e}{\hbar}\right) \label{eq12}
 \end{equation} 
\end{widetext}}
Here $\mathrm{h}_{\mathrm{rf}}$ is the RF magnetic field of $1$ Oe (at 15 dBm rf power) in the strip line of the coplanar wave guide. The calculated values of $\mathrm{J}_{\mathrm{s}}$ are found to be dependent on interlayer (Pt) thickness and vary within the range of 0.135$\pm $0.003 to 0.242$\pm$0.004 MA/m$^{2}$ at 3 GHz as shown in Fig.~\ref{fig5}c.

\section{Conclusion}

This study provides insights into the magnetization dynamics and spin pumping properties at the ML-MoS$_{2}$/Py(5 nm) interface. The distinct two-peak feature in the FMR spectrum is attributed to variations in local magnetization caused by the heterogeneous contact between ML-MoS$_{2}$ and Py, as well as the positioning of MoS$_{2}$ islands. Introducing a Pt interlayer merges these peaks into a single resonance, indicating improved spin transfer efficiency and a more continuous interface. The systematic enhancement in the Gilbert damping factor with increasing Pt interlayer thickness (1, 2, and 5 nm) confirms efficient spin angular momentum transfer from the Py layer to the adjacent material, facilitated by the strong spin orbit coupling (SOC) in Pt. These findings suggest that ML-MoS$_{2}$, in combination with heavy metals such as Pt, is a promising material for spintronic applications, offering potential for effective spin current manipulation in future devices.
\section{Acknowledgements}
M.T. acknowledges the Ministry of Human Resource Development (MHRD), Government of India, for providing a Senior Research Fellowship. S.M. acknowledges the Department of Science and Technology (DST), Government of India, for financial support. R.M. acknowledges financial support from the Initiation Grant, IIT Kanpur (IITK/PHY/2022027), and the I-HUB Quantum Technology Foundation (I-HUB/PHY/2023288), IISER Pune.

\end{document}